\documentclass[pra,twocolumn,superscriptaddress,showpacs,preprintnumbers,amsmath,amssymb]{revtex4}
\usepackage{bm}
\usepackage{bbm}
\usepackage{amssymb}
\usepackage{amsfonts}
\usepackage{epsfig,graphicx}
\usepackage{amstext}
\usepackage{amsmath}
\usepackage{graphicx}
\usepackage{times}
\usepackage{dcolumn}

\begin{document}


\title{Relative quantum coherence, incompatibility, and quantum correlations of states}

\author{Ming-Liang Hu}
\email{mingliang0301@163.com}
\affiliation{Institute of Physics, Chinese Academy of Sciences, Beijing 100190, China}
\affiliation{School of Science, Xi'an University of Posts and Telecommunications, Xi'an 710121, China}
\author{Heng Fan}
\email{hfan@iphy.ac.cn}
\affiliation{Institute of Physics, Chinese Academy of Sciences, Beijing 100190, China}
\affiliation{School of Physical Sciences, University of Chinese Academy of Sciences, Beijing 100190, China}
\affiliation{Collaborative Innovation Center of Quantum Matter, Beijing 100190, China}

\begin{abstract}
Quantum coherence, incompatibility, and quantum correlations are
fundamental features of quantum physics. A unified view of those
features is crucial for revealing quantitatively their intrinsic
connections. We define the relative quantum coherence of two states
as the coherence of one state in the reference basis spanned by the
eigenvectors of another one and establish its quantitative
connections with the extent of mutual incompatibility of two states.
We also show that the proposed relative quantum coherence, which can
take any form of measures such as $l_1$ norm and relative entropy,
can be interpreted as or connected to various quantum correlations
such as quantum discord, symmetric discord, entanglement of
formation, and quantum deficits. Our results reveal conceptual
implications and basic connections of quantum coherence, mutual
incompatibility, and quantum correlations.
\end{abstract}


\maketitle

\section{Introduction}
Quantum coherence is rooted in the superposition of quantum states,
and also remains as a research focus since the early days of quantum
mechanics \cite{Ficek}. It plays a key role for nearly all the novel
quantum phenomena in the fields of quantum optics \cite{Ficek},
quantum thermodynamics \cite{ther1,ther2, ther3,ther4,ther5}, and
quantum biology \cite{biolo}. However, its characterization and
quantification from a mathematically rigorous and physically
meaningful perspective has been achieved only very recently
\cite{coher}, when Baumgratz \emph{et al.} introduced the defining
conditions for a \emph{bona fide} measure of coherence, and proved
that the $l_1$ norm and relative entropy satisfy the required
conditions.

In the past few years, many other coherence measures that satisfy
the defining conditions have been proposed from different aspects
\cite{meas1,meas2,meas3, meas4,meas5,meas6,meas7}. Moreover, the
freezing phenomenon of coherence in open systems \cite{frozen1,
frozen2,frozen3,frozen4,frozen5,frozen6}, the coherence-preserving
channels \cite{preser}, and the creation of coherence by local or
nonlocal operations \cite{creat1,creat2,creat3,creat4} have been
extensively studied. Other topics such as the coherence distillation
\cite{dist1,dist2,dist3}, the complementarity relations of coherence
\cite{comple1,comple2}, the connections of coherence with path
distinguishability \cite{path1,path2} and asymmetry \cite{asym1,
asym2}, the coherence averaged over all basis sets \cite{comple1} or
the Haar distributed pure states \cite{avera}, and the role of
coherence in state merging \cite{qsm} were also studied.

For bipartite and multipartite systems, quantum coherence also
underpins different forms of quantum correlations \cite{qcor,creat2,
distri,meas1,qdcohe,qpt0,qpt1,qpt2,qpt3}. The mutual incompatibility
of states, which represents another fundamental feature of the
nonclassical systems, is also intimately related to quantum
correlations, e.g., for any nondiscordant state there must exist
local measurements which commutes with it \cite{RMP}. Intuitively,
coherence, incompatibility, and quantum correlations are all closely
related concepts. A direct and quantitative connection between them
can provide a whole view and a measure in characterizing the
quantumness of the nonclassical systems. In this paper, by defining
the relative quantum coherence (RQC) of two states as the coherence
of one state in the basis spanned by the eigenvectors of another
one, we find connections between coherence, incompatibility, and
quantum correlations (Fig. \ref{fig:1}). Our observations include:
(i) the quantitative connection between RQC and mutual
incompatibility of states, and (ii) the interpretation of quantum
discord (QD), symmetric discord, measurement-induced disturbance,
quantum deficits, entanglement of formation, and distillable
entanglement via the proposed RQC. We expect the connections
established in this paper may contribute to a unified view of the
resource theory of coherence, incompatibility and quantum
correlations.

\section{Quantifying the RQC}
Consider two states $\rho$ and $\sigma$ in the same Hilbert space
$\mathcal {H}$. When $\sigma$ is nondegenerate with the eigenvectors
$\Xi= \{|\psi_i \rangle\}$, i.e., $\sigma=\sum_i \epsilon_i
|\psi_i\rangle\langle \psi_i|$ ($\epsilon_i$ are the eigenvalues of
$\sigma$), we define the RQC of $\rho$ with respect to $\sigma$ as
\begin{eqnarray}\label{eq2-1}
 C(\rho,\sigma)= C^{\Xi}(\rho),
\end{eqnarray}
where $C^{\Xi}(\rho)$ denotes any \emph{bona fide} measure of
quantum coherence defined in the reference basis $\Xi$.

The rationality for this choice of reference basis lies in that the
RQC of a state with respect to itself equals zero. Although this
definition is essentially the same as that of the coherence measure
introduced by Baumgratz \emph{et al.} \cite{coher}, the choice of
the eigenvectors of another state as the reference basis allows one
to establish quantitatively the connections between quantum
coherence, incompatibility, and quantum correlations of states
(e.g., the excess of RQC of the total state with respect to the
postmeasurement state and the sum of RQC of the reduced states with
respect to the local postmeasurement states gives an interpretation
of the QD), hence can deepen our understanding about distribution of
quantumness in a composite system. Moreover, by choosing
$\sigma=\rho_0$ as the initial state and $\rho=\rho_t$ as the
evolved state, $C(\rho_t,\rho_0)$ also allows one to characterize
the decoherence or coherence process of a system relative to its
initial state.

If one takes the $l_1$ norm or the relative entropy as the coherence
measure \cite{coher}, then
\begin{equation}\label{eq2-2}
 \begin{split}
 &C_{l_1}(\rho,\sigma)=\sum_{i\neq j}|\langle\psi_i|\rho|\psi_j\rangle|,\\
 &C_\mathrm{re}(\rho,\sigma)=S(\Xi|\rho)-S(\rho),
 \end{split}
\end{equation}
where $S(\Xi|\rho)=-\sum_i \langle\psi_i|\rho|\psi_i\rangle\log_2
\langle\psi_i|\rho|\psi_i\rangle$, and $S(\rho)=-\mathrm{tr}
(\rho\log_2 \rho)$ denotes the von Neumann entropy.

When $\sigma$ is degenerate, $\Xi$ is not uniquely defined; however,
we can take the supremum over all possible eigenvectors of $\sigma$
and obtain \emph{the maximum} RQC as
\begin{eqnarray}\label{eq2-3}
 \tilde{C}(\rho,\sigma)=\sup_{\Xi} C^{\Xi}(\rho).
\end{eqnarray}
Although there exists an optimization process, for certain cases it
can be obtained analytically. For example, for the maximally mixed
state $\sigma_{\mathrm m}$, $\tilde{C}(\rho,\sigma_{\mathrm m})$ is
in fact the maximum coherence of $\rho$ over all possible reference
basis. By decomposing $\rho$ (with dimension $d$) via the
orthonormal operator bases $\{X_i\}$ \cite{frozen5}, one can obtain
(see Appendix \ref{sec:5})
\begin{equation}\label{eq2-4}
 \begin{split}
  & \tilde{C}_{l_1}(\rho,\sigma_{\mathrm m})=\sqrt{(d^2-d)/2}|\vec{x}|,\\
  & \tilde{C}_\mathrm{re}(\rho,\sigma_{\mathrm m})=\log_2 d-S(\rho),
 \end{split}
\end{equation}
where $|\vec{x}|$ is the length of the vector $\vec{x}=
(x_1,x_2,\cdots, x_{d^2-1})$, and $x_i=\mathrm{tr}(\rho X_i)$. As
$|\vec{x}|^2 \leqslant 2(d-1)/d$ and $S(\rho) \geqslant 0$
\cite{frozen5}, we have $\tilde{C}_{l_1}(\rho,\sigma_{\mathrm
m})\leqslant d-1$ and $\tilde{C}_{\rm re}(\rho,\sigma_{\mathrm m})
\leqslant \log_2 d$, and the bounds are achieved for pure states
$\rho$.

Equation \eqref{eq2-4} reveals that the states without quantum
coherence with respect to any basis are the maximally mixed states
and the states with maximum coherence with respect to any basis are
the pure states. This property is similar to the quantumness
captured by the non-commutativity of the algebra of observables
$\mathcal {A}$, in which a state $\rho$ is defined to be classical
if and only if $\mathrm{tr}(\rho[A,B])=0$, $\forall A,B\in \mathcal
{A}$ \cite{nonc}. Here, it is easy to check that only
$\sigma_{\mathrm m}$ is classical in this sense of the definition.
Moreover, the maximal coherence of Eq. \eqref{eq2-4} is also
intimately related to the complementarity of coherence under
mutually unbiased bases (MUBs). For example, from Eq. (13) of Ref.
\cite{comple1} one can obtain that the upper bound for the sum of
the squared $l_1$ norm of coherence is just $dC_{l_1}^2
(\rho,\sigma_{\mathrm m})$, i.e., $\sum_{j=1}^{d+1} C_{l_1}^2
(A_j,\rho)\leqslant dC_{l_1}^2 (\rho,\sigma_{\mathrm m})$, with
$\{A_j\}_{j=1}^{d+1}$ being the MUBs, and for the one-qubit state
the equality always holds. Similarly, from Eq. (23) of
\cite{comple1} we obtain $\sum_{j=1}^{d+1} C_\mathrm{re}(A_j,\rho)
\leqslant (d+1)C_\mathrm{re}(\rho,\sigma_{\mathrm m})- C_{l_1}^2
(\rho,\sigma_{\mathrm m})\log_2(d-1)/[d(d-2)]$. This gives an upper
bound for the sum of the relative entropy of coherence under MUBs
via the difference of maximal RQCs measured by the relative entropy
and the $l_1$ norm.

The RQC is unitary invariant in the sense that $C_{l_1}(\rho,\sigma)
= C_{l_1}(U\rho U^\dag, U\sigma U^\dag)$ for any unitary operation
$U$. This can be proved directly by noting that $U\sigma
U^\dag=\sum_i \epsilon_i |\psi_i^U \rangle \langle \psi_i^U|$ and
$U\rho U^\dag=\sum_{kl}\lambda_{kl} |\psi_k^U\rangle \langle
\psi_l^U|$, with $|\psi_i^U\rangle = U|\psi_i\rangle$. That is, $U$
rotates both the basis of $\rho$ and $\sigma$ simultaneously.

\section{Linking RQC to incompatibility of states}
We establish connections between the RQC and mutual incompatibility
of two states in this section. For this purpose, we rewrite $\rho$
in the basis $\Xi$ as $\rho= \sum_{kl}\lambda_{kl} |\psi_k\rangle
\langle \psi_l|$, where $\lambda_{kl}=\langle\psi_k |\rho|\psi_l
\rangle$. Then by Eq. \eqref{eq2-2} we obtain
\begin{equation}\label{eq2-5}
 C_{l_1}(\rho,\sigma)=\sum_{i\neq j}|\lambda_{ij}|.
\end{equation}
On the other hand, the commutator $[\rho,\sigma]$ of states $\rho$
and $\sigma$ can be calculated as
\begin{equation}\label{eq2-6}
 [\rho,\sigma]=\sum_{i\neq j}\lambda_{ij}(\epsilon_j-\epsilon_i)
               |\psi_i\rangle\langle\psi_j|,
\end{equation}
then if we quantify the extent of the mutual incompatibility of the
two states by $l_1$ norm of the commutator, we have
\begin{equation}\label{eq2-7}
  Q_{l_1}(\rho,\sigma) =\sum_{i\neq j}|\lambda_{ij}(\epsilon_i-\epsilon_j)|,
\end{equation}
where $Q_{l_1}(\rho,\sigma)$ is by definition nonnegative, vanishes
if and only if the commutator $[\rho, \sigma]$ vanishes, and is
unitary invariant. Different from the mutual incompatibility
$Q_F(\rho,\sigma)=2\|[\rho,\sigma]\|_2^2$ measured by square of the
Frobenius norm, which takes the value between 0 and 1 \cite{nonc},
the maximum possible value of $Q_{l_1}(\rho,\sigma)$ is
$\sqrt{d-1}$. But apart from the single-qubit case, this maximum is
reached on the pure but not maximally coherent states (see Appendix
\ref{sec:6}).

Here, by comparing Eqs. \eqref{eq2-5} and \eqref{eq2-7}, one can
obtain a quantitative connection between $C_{l_1}(\rho,\sigma)$ and
$Q_{l_1}(\rho,\sigma)$,
\begin{equation}\label{eq2-8}
  C_{l_1}(\rho,\sigma)\geqslant Q_{l_1}(\rho,\sigma),
\end{equation}
due to $|\epsilon_i-\epsilon_j|\leqslant 1$, $\forall~ i,j$. For the
special case that $\sigma$ is pure, it has only one nonvanishing
eigenvalue; this gives $Q_{l_1}(\rho,\sigma) = 2\sum_{i\neq
1}|\lambda_{i1}|$. Thus when the elements $\lambda_{kl}$ of $\rho$
not in the first row, the first column, and the main diagonal equal
to zero, we have $C_{l_1}(\rho,\sigma) = Q_{l_1} (\rho,\sigma)$.
That is to say, the bound in Eq. \eqref{eq2-8} is tight. In
particular, for pure state $\sigma$ and arbitrary state $\rho$ of
one qubit, the bound is always saturated.

\begin{figure}
\centering
\resizebox{0.43 \textwidth}{!}{%
\includegraphics{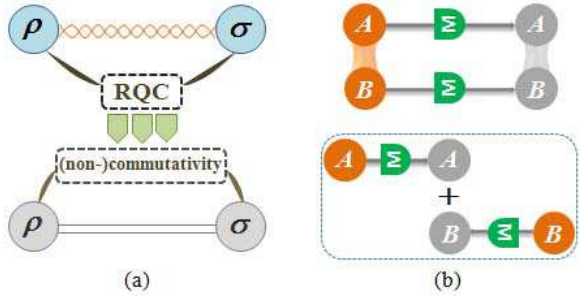}}
\caption{(Color online) (a) Equivalence between RQC and
commutativity of states. The vanishing (nonvanishing) RQC implies
commutativity (non-commutativity) of states and vice versa. (b)
Interpretations of quantum correlations via RQC. The various
correlations can be interpreted as or connected to the discrepancy
between RQC for the total state (top) and those localized in the
reduced states (bottom).}\label{fig:1}
\end{figure}

From Eq. \eqref{eq2-8}, one can see that the mutual incompatibility
of two states provides a lower bound for their RQC. Alternatively,
the RQC provides an upper bound for the extent of their mutual
incompatibility. It also implies that when there is no RQC between
two states, then they must commute with each other. On the other
hand, when two states commute, they either share a common eigenbasis
or are orthogonal, for which we always have the vanishing RQC. What
is more, by using Eqs. \eqref{eq2-6}, \eqref{eq2-7}, and the
definition of $Q_F (\rho,\sigma)$ \cite{nonc}, one can show that
$Q_{l_1}^2(\rho,\sigma) \geqslant Q_{F}(\rho,\sigma)/2 $. As
$Q_{F}(\rho,\sigma)$ can be measured via an interferometric setup
and without performing the full state tomography \cite{nonc}, it
provides an experimentally accessible lower bound for the mutual
incompatibility and RQC of two quantum states.

In fact, for every basis state $|\psi_i\rangle$ of $\sigma$, we can
associate with it a pure state $\sigma_i= |\psi_i\rangle\langle
\psi_i|$. Then by summing $Q_{l_1}(\rho,\sigma_i)$ over the set
$\{\sigma_i\}$, one can obtain
\begin{equation}\label{eq2-9}
 \sum_i Q_{l_1}(\rho,\sigma_i)= 2C_{l_1}(\rho,\sigma),
\end{equation}
that is to say, the sum of $Q_{l_1}(\rho,\sigma_i)$ over
$\{\sigma_i\}$ equals twice the RQC between $\rho$ and $\sigma$.
This establishes another connection between RQC and the mutual
incompatibility of two states.

\section{Linking RQC to quantum correlations}
The RQC and incompatibility of states are also intimately related to
quantum correlations such as QD \cite{QD}. For bipartite state
$\rho_{AB}$ with the reduced state $\rho_A$, if the QD
$D_A(\rho_{AB})$ defined with respect to party $A$ equals zero, then
$\rho_{AB}$ and $\rho_A\otimes \mathbb{I}_B$ commute. Equivalently,
if $\rho_{AB}$ does not commute with $\rho_A\otimes \mathbb{I}_B$,
then it is quantum discordant \cite{nullqd}. By Eq. \eqref{eq2-8},
we know that the nonvanishing mutual incompatibility implies the
nonvanishing RQC of $\rho_{AB}$ with respect to $\rho_A \otimes
\mathbb{I}_B$. In fact, a necessary and sufficient condition for
$\rho_{AB}$ to have zero discord has also been proved, which says
that $D_A(\rho_{AB})=0$ if and only if all the operators
$\rho_{A|b_1 b_2}= \langle b_1 |\rho_{AB} |b_2 \rangle$ commute with
each other for any orthonormal basis $\{|b_i\rangle\}$ in
$\mathcal{H}_B$ \cite{nullqd2}. From the analysis below Eq.
\eqref{eq2-8}, we know that the commutativity of two operators
corresponds to the vanishing RQC of them. Hence, the RQC can be
linked to QD of a state.

By denoting $\mathbb{P}=\{\Pi_k^A\}$ ($\Pi_k^A= |k\rangle \langle
k|$) the local projective measurements on party $A$, and likewise
for $\mathbb{Q}=\{\Pi_l^B\}$, the postmeasurement states after the
measurements $\mathbb{P}\otimes \mathbb{I}_B$ and
$\mathbb{P}\otimes\mathbb{Q}$ are given, respectively, by
\begin{equation}\label{eq3-1}
 \begin{aligned}
  & \rho_{\mathbb{P}B}=\sum_k p_k \Pi_k^A\otimes \rho_{B|k},\\
  & \rho_{\mathbb{PQ}}=\sum_{kl} p_{kl}\Pi_k^A\otimes \Pi_l^B,
 \end{aligned}
\end{equation}
where $\rho_{B|k}= \mathrm{tr}_A (\Pi_k^A\rho_{AB}\Pi_k^A)/p_k$,
$p_k=\langle k| \rho_A| k\rangle$, $p_{kl}=\langle k l |\rho_{AB}|k
l \rangle$. Then, by the definition of QD \cite{QD}, we obtain
\begin{equation}\label{eq3-2}
 \begin{split}
  D_A(\rho_{AB})&= S(\rho_A)-S(\rho_{AB})-S(\tilde{\rho}_{\mathbb{P}})+S(\tilde{\rho}_{\mathbb{P}B})\\
        & \leqslant S(\rho_A)-S(\rho_{AB})-S(\tilde{\rho}_{\mathbb{P}})+S(\tilde{\rho}_{\mathbb{PQ}})\\
        & =C_{\rm re}(\rho_{AB},\tilde{\rho}_{\mathbb{PQ}})
          -C_{\rm re}(\rho_A,\tilde{\rho}_\mathbb{P})\\
        & \equiv\delta_1(\rho_{AB}),
 \end{split}
\end{equation}
where $S(\tilde{\rho}_{\mathbb{P}B}) \leqslant
S(\tilde{\rho}_{\mathbb{PQ}})$ as project measurements do not
decrease entropy. Moreover, $\tilde{\rho}_{\mathbb{P}B}$ and
$\tilde{\rho}_{\mathbb{PQ}}$ denote, respectively, the
postmeasurement states of $\{\tilde{\Pi}_k^A\}$ and
$\{\tilde{\Pi}_k^A \otimes\tilde{\Pi}_l^B\}$, where
$\{\tilde{\Pi}_k^A\}$ are the optimal measurements for obtaining
$D_A(\rho_{AB})$, while $\{\tilde{\Pi}_l^B\}$ can in fact be
arbitrary projective measurements, and here we fix it to be the
measurement that gives the minimal entropy increase of
$\tilde{\rho}_{\mathbb{P}B}$ for tightening the above bound.
Moreover, the eigenbasis of $\tilde{\rho}_{\mathbb{PQ}}$
($\tilde{\rho}_{\mathbb{P}}$) for obtaining the RQC in Eq.
\eqref{eq3-2} are chosen to be $\{|\tilde{k}\rangle
\otimes|\tilde{l}\rangle\}$ ($\{|\tilde{k}\rangle\}$), which
corresponds to the optimal measurement $\{\tilde{\Pi}_k^A \otimes
\tilde{\Pi}_l^B\}$, i.e., here we do not perform the optimization of
Eq. \eqref{eq2-3} even if the postmeasurement states are degenerate.
The same holds for other discussions of this section.

In Eq. \eqref{eq3-2}, $C_{\rm re} (\rho_{AB},
\tilde{\rho}_{\mathbb{PQ}})$ is the RQC (defined by the relative
entropy) between $\rho_{AB}$ and $\tilde{\rho}_{\mathbb{PQ}}$, while
$C_{\rm re}(\rho_A,\tilde{\rho}_\mathbb{P})$ is that between
$\rho_A$ and $\tilde{\rho}_\mathbb{P}$, which can be recognized as
the quantum coherence localized in subsystem $A$. From this point of
view, the QD of a state is always smaller than or equal to
$\delta_1(\rho_{AB})$, which characterizes the discrepancy between
the RQCs for the total system and that for the subsystem to be
measured in the definition of QD, see Fig. \ref{fig:1}(b). When the
RQC discrepancy $\delta_1(\rho_{AB})$ vanishes, there will be no QD
in the state.

For the case of quantum-classical state $\chi_{AB}=\sum_l p_l
\rho_{A|l} \otimes |\varphi_l\rangle \langle\varphi_l|$, with
$\rho_{A|l}$ being the density operator in $\mathcal {H}_A$,
$|\varphi_l \rangle$ the orthonormal basis (also the eigenvectors of
$\chi_B=\mathrm{tr}_A \chi_{AB}$) in $\mathcal {H}_B$, and $\{p_l\}$
the probability distribution, we have $\tilde{\rho}_{\mathbb{P}B}=
\tilde{\rho}_{\mathbb{PQ}}$ if we choose $\mathbb{Q}= \{|\varphi_l
\rangle \langle\varphi_l|\}$. Then the RQC discrepancy
$\delta_1(\chi_{AB})$ equals exactly the QD of $\chi_{AB}$, i.e.,
\begin{equation}
 D_A(\chi_{AB})= C_{\rm re}(\chi_{AB},\tilde{\chi}_{\mathbb{PQ}})
        -C_{\rm re}(\chi_A,\tilde{\chi}_\mathbb{P}),
\end{equation}
and hence the bound given in Eq. \eqref{eq3-2} is tight.

Besides the QD defined with one-sided measurements \cite{QD}, one
can also demonstrate the role of RQC discrepancy in interpreting the
symmetric discord $D_{s}(\rho_{AB})= I(\rho_{AB})-
I(\tilde{\rho}_\mathbb{PQ})$ defined via two-sided optimal
measurements $\{\tilde{\Pi}_k^A \otimes \tilde{\Pi}_l^B\}$
\cite{sqd}. For this case, one can prove that
\begin{equation}\label{eq3-9}
 \begin{split}
 D_{s}(\rho_{AB}) =& C_{\rm re}(\rho_{AB},\tilde{\rho}_\mathbb{PQ})
                     -C_{\rm re}(\rho_A,\tilde{\rho}_\mathbb{P})\\
                   &-C_{\rm re}(\rho_B,\tilde{\rho}_\mathbb{Q})
                    \equiv \delta_2 (\rho_{AB}),
 \end{split}
\end{equation}
with  $\delta_2(\rho_{AB})$ being the RQC discrepancy. It implies
that the symmetric discord $D_{s}(\rho_{AB})$ is nonzero if and only
if there exists RQC not localized in the subsystems, see Fig.
\ref{fig:1}(b). This establishes a direct connection between the RQC
discrepancy and the symmetric discord.

The RQC can also be linked to other discord-like correlation
measures such as measurement-induced disturbance \cite{MID},
measurement-induced nonlocality \cite{MIN}, and quantum deficits
\cite{defi1,defi2}. (i) For the measurement-induced disturbance, it
is just the RQC of $\rho_{AB}$ with respect to $\rho'_{AB}=
\sum_{ij} \xi_{ij} \rho \xi_{ij}$, i.e., $M(\rho_{AB})=C_\mathrm{re}
(\rho_{AB}, \rho'_{AB})$ \cite{hufan}. Here $\xi_{ij}= |e_i^A\rangle
\langle e_i^A|\otimes |e_j^B\rangle \langle e_j^B|$, and
$\{|e_i^A\rangle, |e_j^B\rangle \}$ are local eigenvectors of the
reduced states. (ii) For the measurement-induced nonlocality defined
as $N_v(\rho_{AB})= \max_{\Pi^A} S(\Pi^A[\rho])-S(\rho)$ ($\Pi^A$
are restricted to the locally invariant measurements), from Eq. (9)
of Ref. \cite{MIN} one can obtain that $N_v(\rho_{AB}) \leqslant
C_\mathrm{re}(\rho_{AB}, \tilde{\rho}_\mathbb{PQ})$. Here,
$\tilde{\rho}_\mathbb{PQ}$ is similar to that in Eq. \eqref{eq3-2},
and the difference is that $\{\tilde{\Pi}_k^A\}$ should be locally
invariant. (iii) For the zero-way deficit $\Delta^{\varnothing}=
S(\tilde{\rho}_ \mathbb{PQ})- S(\rho_{AB})$ and one-way deficit
(equal to the thermal discord \cite{RMP}) $\Delta^{\rightarrow}
=S(\tilde{\rho}_{\mathbb{P}B})-S(\rho_{AB})$ (where $\tilde{\rho}_
\mathbb{PQ}$ and $\tilde{\rho}_{\mathbb{P}B}$ denote, respectively,
the corresponding optimal postmeasurement states) \cite{defi2}, it
is direct to see that
\begin{equation}\label{eq3-10}
 \begin{split}
  & \Delta^{\varnothing}(\rho_{AB})=C_{\rm re}(\rho_{AB},\tilde{\rho}_\mathbb{PQ}),\\
  & \Delta^{\rightarrow}(\rho_{AB})\leqslant C_{\rm re}(\rho_{AB},\tilde{\rho}_\mathbb{PQ}),
 \end{split}
\end{equation}
where for the quantum-classical states, the inequality becomes
equality when $\mathbb{Q}= \{|\varphi_l\rangle \langle\varphi_l|\}$.
These relations give interpretations of the corresponding
correlation measures in terms of RQC, and hence bridge the gap
between quantum coherence for a single quantum system and quantum
correlations for a system with two parties, which are two
fundamentals of quantum physics.

For certain cases, the RQC can also be linked to quantum
entanglement. For example, by resorting to the Koashi-Winter
equality \cite{Kwinter}, the chain inequality \cite{chain}, and
equality condition of the Araki-Lieb inequality \cite{xizj}, one can
obtain
\begin{equation}\label{eq3-5}
  E_f(\rho_{AB})\leqslant \delta_1(\rho_{AB}),
\end{equation}
when the conditional entropy $S(B|A)$ is nonnegative or when the
equality $S(\rho_{AB})=|S(\rho_A)-S(\rho_B)|$ is satisfied (see
Appendix \ref{sec:7}). Here, $E_f (\rho_{AB})$ denotes the
entanglement of formation for $\rho_{AB}$ \cite{EoF}. The relation
shows that for these cases, the entanglement of formation for a
state is always bounded from above by its RQC discrepancy
$\delta_1(\rho_{AB})$.

When one performs local measurements $\{\Pi_k^A\}$ on a system
$\rho_{AB}$, there will be entanglement created between the
measurement apparatus $M$ and the system $AB$. Streltsov \emph{et
al.} showed that the minimal distillable entanglement
$E_D^{\min}(\tilde{\rho}_{M:AB})= \min_{U} E_D(\tilde{\rho}_{M:AB})$
created between $M$ and $AB$ equals $\Delta^{\rightarrow}
(\rho_{AB})$ \cite{new1}. Here, $\tilde{\rho}_{MAB}= U(|0_M\rangle
\langle 0_M| \otimes \rho_{AB}) U^\dag$, and $U$ are unitaries
acting on $MAB$ which realize a von Neumann measurement on $A$,
i.e., $\mathrm{tr}_M \tilde{\rho}_{MAB}=\sum_k \Pi_k^A
\rho_{AB}\Pi_k^A$. Then by Eq. \eqref{eq3-10} we obtain $E_D^{\min}
(\tilde{\rho}_{M:AB}) \leqslant C_\mathrm{re}(\rho_{AB},
\tilde{\rho}_\mathbb{PQ})$. Moreover, the minimal partial
entanglement $P_{E_D}^{\min} (\tilde{\rho}_{MAB}) =\min_U
[E_D(\tilde{\rho}_{M:AB})- E_D(\tilde{\rho}_{MA})]$ created in a von
Neumann measurement on $A$ equals $D_A(\rho_{AB})$ \cite{new1}, and
this gives $P_{E_D}^{\min}(\tilde{\rho}_{M:AB})\leqslant
\delta_1(\rho_{AB})$. All these relations show the role of RQC in
interpreting quantum entanglement.

\section{Conclusion}
In this paper, we explored the connections among quantum coherence,
incompatibility, and quantum correlations. We defined the RQC and
proved several of its connections to the extent of mutual
incompatibility of states. We also gave interpretations of various
quantum correlations (QD, entanglement of formation, symmetric
discord, measurement-induced disturbance, measurement-induced
nonlocality, and quantum deficits, etc.) via the RQC discrepancy
between the total system and those localized in the respective
subsystems.

We mainly considered the $l_1$ norm and relative entropy of
coherence, but we remark that the coherence measure in other forms
can also be extended to the case of RQC, and more connections among
coherence, mutual incompatibility, and quantum correlations can be
expected in future research.

\emph{Noted added.} Recently Yao \emph{et al.} presented a similar
study on the problem of maximum coherence under generic basis
\cite{added}.

\section*{ACKNOWLEDGMENTS}
This work was supported by NSFC (Grants No.
11675129 and 91536108), MOST (Grants No. 2016YFA0302104 and
2016YFA0300600), New Star Project of Science and Technology of
Shaanxi Province (Grant No. 2016KJXX-27), CAS (Grants No.
XDB01010000 and XDB21030300), and New Star Team of XUPT.

\begin{appendix}
\section{Proof of Eq. (4)}\label{sec:5}
\setcounter{equation}{0}
\renewcommand{\theequation}{A\arabic{equation}}
\setcounter{figure}{0}
\renewcommand{\thefigure}{A\arabic{figure}}

For any $d$-dimensional state $\rho$, it can always be decomposed as
\begin{equation}\label{eqa-00}
 \rho = \frac{1}{d}\mathbb{I}_d+\frac{1}{2}\sum_{i=1}^{d^2-1} x_iX_i,
\end{equation}
where $\{X_i\}$ constitutes the orthonormal operator bases, and the
$l_1$ norm of coherence can be obtained as \cite{comple2}
\begin{equation}\label{eqa-01}
 C_{l_1}(\rho)=\sum_{r=1}^{d_0}(x_{2r-1}^2+ x_{2r}^2)^{1/2},
\end{equation}
where $d_0=d(d-1)/2$.

The maximization of $C_{l_1}(\rho)$ over all possible reference
basis is equivalent to its maximization over all the unitary
transformations of $\rho$ which keep $|\vec{x}|$ unchanged. Then, by
denoting $\tilde{x}_i={\rm tr}(\rho^U X_i)$ with $\rho^U= U\rho
U^\dag$, and by using the mean inequality, we obtain
\begin{equation}\label{eqa-1}
 \begin{split}
  C_{l_1}(\rho^U)&=\sum\nolimits_{r=1}^{d_0}\sqrt{\tilde{x}_{2r-1}^2+\tilde{x}_{2r}^2}\\
                 & \leqslant \sqrt{d_0 \sum\nolimits_{k=1}^{2d_0}\tilde{x}_k^2}\\
                 & \leqslant \sqrt{d_0} |\vec{x}|,
 \end{split}
\end{equation}
and $C_{l_1}(\rho^U)=\sqrt{d_0} |\vec{x}|$ when $U$ gives rise to
$\tilde{x}_{2r-1}^2+\tilde{x}_{2r}^2= \tilde{x}_{2r'-1}^2+
\tilde{x}_{2r'}^2$, $\forall~ r,r'\in[1,d_0]$, and $\tilde{x}_l=0$,
$\forall~ l\geqslant 2d_0+1$. Thus we arrive at the first equality
of Eq. (4).

Second, the maximization of $C_\mathrm{re}(\rho,\sigma_\mathrm{m})$
over all the reference basis is equivalent to the maximization of
$S(\rho_\mathrm{diag}^U)$ over all $U$. Still, by using the mean
inequality, we obtain
\begin{equation}\label{eqa-2}
 \begin{split}
  S(\rho_\mathrm{diag}^U)&=-\sum\nolimits_{i=1}^d\rho^U_{ii}\log_2\rho^U_{ii}\\
                      &\leqslant \sqrt{d\sum\nolimits_{i=1}^d (\rho^U_{ii}\log_2\rho^U_{ii})^2}\\
                      &\leqslant \log_2 d,
 \end{split}
\end{equation}
where $S(\rho_\mathrm{diag}^U)=\log_2 d$ when $\rho^U_{ii} \log_2
\rho^U_{ii}= \rho^U_{jj}\log_2\rho^U_{jj}$, $\forall~ i,j$, and
$\tilde{x}_l=0$, $\forall~ l\geqslant 2d_0+1$. This completes the
proof of the second equality of Eq. (4).

\section{Maximum of the mutual incompatibility}\label{sec:6}
\setcounter{equation}{0}
\renewcommand{\theequation}{B\arabic{equation}}

To obtain the maximum of $Q_{l_1}(\rho,\sigma)$, we consider
$\rho^{\Psi}=|\Psi\rangle\langle\Psi|$, with $|\Psi\rangle=
\sum_{i=1}^d a_i^{(d)} |\psi_i\rangle$ and $\sum_{i=1}^d |a_i^{(d)}|
=1$, and $\sigma_1=|\psi_1\rangle\langle\psi_1|$, for which we have
\begin{equation}\label{eqb-1}
 Q_{l_1}(\rho^\Psi,\sigma_1)=2\sum_{i\neq 1}|a_i^{(d)}a_1^{(d)}|.
\end{equation}
Then by taking $a_1^{(2)}=\cos\theta_2$ and $a_2^{(2)}= \sin
\theta_2$ for $d=2$, and $a_1^{(d)}=\cos\theta_d$, $a_j^{(d)}=
\sin\theta_d a_{j-1}^{(d-1)}$ for $j\geqslant 2$ and $d\geqslant 3$
(the phases of $a_i^{(d)}$ do not affect the incompatibility), one
can show
\begin{equation}\label{eqb-2}
 Q_{l_1}^{\rm max}(\rho^\Psi, \sigma_1)=\sqrt{d-1},
\end{equation}
which is obtained with $\theta_2=\theta_d=(2n+1)\pi/4$ $(n\in
\mathbb{Z})$, and $\theta_k=\pi/2-\arctan{(1/\sqrt{k-1})}$,
$\forall~ k\in [3,d-1]$.

Now, we further prove that $\sqrt{d-1}$ is also the maximum of
$Q_{l_1}(\rho,\sigma)$ for general cases. First, for the pure states
$\rho^\Psi$ and general $\sigma$, the convexity of the $l_1$ norm
gives
\begin{equation}\label{eqb-3}
 \begin{split}
 Q_{l_1}(\rho^\Psi,\sigma)&\leqslant \sum\nolimits_i \epsilon_i Q_{l_1}(\rho^\Psi,\sigma_i) \\
                  &\leqslant \sum\nolimits_i \epsilon_i Q_{l_1}(\tilde{\rho}^\Psi,\sigma_1)\\
                  & =\sqrt{d-1},
\end{split}
\end{equation}
where $\tilde{\rho}^\Psi$ is the optimal state for obtaining
$Q_{l_1}^{\rm max}(\rho^\Psi, \sigma_1)$, and the second inequality
comes from the fact that $\tilde{\rho}^\Psi$ is not necessarily the
optimal state for $Q_{l_1}^{\rm max}(\rho^\Psi, \sigma_i)$ when
$i\geqslant 2$.

Second, by denoting $\rho=\sum_i p_i |\phi_i\rangle\langle \phi_i|$
any pure state decomposition of the general state $\rho$, we have
\begin{equation}\label{eqb-4}
 \begin{split}
 Q_{l_1}(\rho,\sigma) &\leqslant \sum\nolimits_i p_i Q_{l_1}(|\phi_i\rangle
                                \langle\phi_i|,\sigma)\\
                      &\leqslant \sqrt{d-1},
 \end{split}
\end{equation}
and the second inequality is due to $Q_{l_1}(|\phi_i\rangle\langle
\phi_i|,\sigma) \leqslant \sqrt{d-1}$ for any $|\phi_i\rangle$.
Therefore, we proved $Q_{l_1}^{\rm max} (\rho,\sigma)=\sqrt{d-1}$.

\section{Proof of Eq. (15)}\label{sec:7}
\setcounter{equation}{0}
\renewcommand{\theequation}{C\arabic{equation}}

First, $S(B|A)\geqslant 0$ implies $S(\rho_C)\geqslant S(\rho_A)$
for any pure state $|\Psi\rangle_{ABC}$. Then, by using the chain
inequality \cite{chain}, one can obtain
\begin{equation}\label{eqc-1}
  S(\rho_C)+E_f(\rho_{AB})\leqslant S(\rho_A)+E_f(\rho_{BC}),
\end{equation}
which is equivalent to $E_f(\rho_{AB})\leqslant E_f(\rho_{BC})-
S(B|A)$. This, together with Eq. \eqref{eq3-2} and the Koashi-Winter
equality \cite{Kwinter}
\begin{equation}\label{eqc-2}
  D_A(\rho_{AB})+S(B|A)= E_f(\rho_{BC}),
\end{equation}
gives $E_f(\rho_{AB})\leqslant \delta_1(\rho_{AB})$.

Second, the equality $S(\rho_{AB})=S(\rho_B)-S(\rho_A)$ if and only
if the Hilbert space $\mathcal{H}_B$ can be decomposed as
$\mathcal{H}_B= \mathcal {H}_{B^L}\otimes \mathcal {H}_{B^R}$ such
that $\rho_{AB}= |\psi\rangle_{AB^L}\langle\psi|\otimes \rho_{B^R}$
\cite{xizj}. For this case, it is direct to show that
\begin{equation}\label{eqc-3}
 E_f(\rho_{AB})= D_A(\rho_{AB})=D_B(\rho_{AB})=S(\rho_A).
\end{equation}
This, together with Eq. \eqref{eq3-2}, gives
$E_f(\rho_{AB})\leqslant \delta_1(\rho_{AB})$.

Similarly, $S(\rho_{AB})=S(\rho_A)-S(\rho_B)$ if and only if there
exists a decomposition $\mathcal{H}_A= \mathcal {H}_{A^L}\otimes
\mathcal {H}_{A^R}$ such that $\rho_{AB}= \rho_{A^L}\otimes
|\psi\rangle_{A^R B}\langle\psi|$, and for this kind of states we
have
\begin{equation}\label{eqc-4}
  E_f(\rho_{AB})=D_A(\rho_{AB})= D_B(\rho_{AB})=S(\rho_B).
\end{equation}
Hence, we still have $E_f(\rho_{AB})\leqslant \delta_1(\rho_{AB})$.

\end{appendix}

\newcommand{\PRL}{Phys. Rev. Lett. }
\newcommand{\RMP}{Rev. Mod. Phys. }
\newcommand{\PRA}{Phys. Rev. A }
\newcommand{\PRB}{Phys. Rev. B }
\newcommand{\PRE}{Phys. Rev. E }
\newcommand{\PRX}{Phys. Rev. X }
\newcommand{\NJP}{New J. Phys. }
\newcommand{\JPA}{J. Phys. A }
\newcommand{\JPB}{J. Phys. B }
\newcommand{\PLA}{Phys. Lett. A }
\newcommand{\NP}{Nat. Phys. }
\newcommand{\NC}{Nat. Commun. }
\newcommand{\SR}{Sci. Rep. }
\newcommand{\EPJD}{Eur. Phys. J. D }
\newcommand{\QIP}{Quantum Inf. Process. }
\newcommand{\AoP}{Ann. Phys. (NY) }
\newcommand{\PR}{Phys. Rep. }
%
%

\end{document}